\documentclass[10pt]{article}
\usepackage{amssymb,amsmath}
\begin{document}

\begin{center}
{\it Proceedings of Quantum Optics V, Cozumel (2010)\\ to appear in Revista Mexicana de Fisica} 
\vspace*{.5cm}

%\title[]{Clifford group dipoles and the enactment of Weyl/Coxeter group $W(E_8)$ by entangling gates}
%\author{Michel Planat}
%\address{Institut FEMTO-ST, CNRS, 32 Avenue de
%l'Observatoire,\\ F-25044 Besan\c con, France (planat@femto-st.fr)}
{\bf  About the Dedekind psi function in Pauli graphs}
\vspace*{.2cm}

Michel Planat

Institut FEMTO-ST, CNRS, 32 Avenue de l'Observatoire,\\ F-25044 Besan\c con, France (planat@femto-st.fr)
\end{center}
%\address{$^{\ddag}$ CNRS I3S, Les Algorithmes, Euclide B,
%2000 route des Lucioles,\\ BP 121, 06903 Sophia Antipolis, France}
 
%\address{$^*$ Department of Computer Science, University of Bristol, Merchant Venturers Building, Bristol, BS8 1TW, U.K.}

\begin{abstract}
We study the commutation structure within the Pauli groups built on all decompositions of a given Hilbert space dimension $q$, containing a square, into its factors. The simplest illustrative examples
are the quartit ($q=4$) and two-qubit ($q=2^2$) systems. It is shown how the sum of divisor function $\sigma(q)$ and the Dedekind psi function $\psi(q)=q \prod_{p|q} (1+1/p)$ enter into the theory for counting 
the number of maximal commuting sets of the qudit system. In the case of a multiple qudit system (with $q=p^m$ and $p$ a prime), the arithmetical functions $\sigma(p^{2n-1})$ and $\psi(p^{2n-1})$ count the cardinality of the symplectic polar space $W_{2n-1}(p)$ that endows the commutation structure and its punctured counterpart, respectively. Symmetry properties of the Pauli graphs attached to these structures are investigated in detail and several illustrative examples are provided.
\end{abstract}

 \begin{center}
Pacs: 03.67.Lx, 02.10.Ox, 02.20.-a, 02.10.De, 02.40.Dr
\end{center}  
    
\section{Introduction}

The paper contains a set of graph, group and number theoretical tools useful to understand the commutation structure of operators in qudit and multiple qudit systems. They are intended to help the development of new quantum algorithms and the design of efficient quantum information systems. 
%For instance, a four-level system, also known as a quartit, and its relation to a two-qubit system is useful to understand in every detail.
For instance, states of nuclear spin $\frac{3}{2}$ (a quartit) in a specific GaAs quantum well device may be used for realizing the logical single and two-qubit gates by applying selective radio frequency pulses at the resonance frequency between two energy levels \cite{Hirayama2006}. Note that a $q$-level system (or qudit), in the Hilbert space dimension $q=\prod_i p_{i}^{\alpha_i}$ written as a product of primes, is equivalent to a mixture of $p_i$ dits when there is no square, i.e. $\alpha_i=1$ for any $i$. The general case is tricky but can be reduced to elementary building blocks using the suitable algebraic tools explained in this paper. In the simplest case, a quartit system ($q=4$) is not a two-qubit sytem ($q=2^2$). This can be seen with the maximal commuting sets. We already emphasized that the geometry of the two-qubit system (with points as the $15$ observables and lines as the $15$ triple of mutually commuting sets) is the (self-dual) generalized quadrangle $GQ(2,2)$, with automorphism group the symmetric group $S_6$ \cite{Planat2007}. Besides, the quartit system contains $15$ observables and $7=6+1$ maximum cliques in the Pauli graph: the decomposition reflects the distinction between the sum of divisor function $\sigma(4)=7$, that counts the whole number of maximum cliques, and the Dedekind psi function $\psi(4)=6$, that counts the maximum cliques belonging to the projective line $\mathbb{P}_1(\mathbb{Z}_4)$. Both functions $\sigma(q)$ and $\psi(q)$ are equal if $q$ does not contain a square \cite{PlanatJPA}.

In this paper, we are interested in the commutation relations of observables attached to a selected decomposition of the Hilbert space dimension $q$. The observables in a factor are defined from the action on a vector $\left|s\right\rangle$ of the $q_i$-dimensional Hilbert space of the $q_i$-dit Pauli group generated by two unitary $X$ (shift) and clock $Z$ operators via $X\left|s\right\rangle=\left|s+1\right\rangle$ and $Z\left|s\right\rangle=\omega^s \left|s\right\rangle$, with $\omega$ a primitive $q_i$-th root of unity. Then the observables in dimension $q$ are obtained by taking tensor products over the $q_i$-dimensional observable of each factor. A Pauli graph is constructed by taking the observables as vertices and a edge joining two commuting observables. Maximal sets of mutually commuting observables, i.e. maximum cliques of the Pauli graph, are used to define a point/line incidence geometry with observables as points and maximum cliques as lines. 

We focus on quantum systems of Pauli observables defined over the Hilbert space of dimension $q$ containing a square. In the single qudit case, studied in Sec. 2, the maximal mutually commuting sets of observables in the Hilbert space of dimension $q$ are mapped bijectively to the maximal submodules over the ring $\mathbb{Z}_q$ \cite{Havlicek2007,Havlicek2008}. If $q$ contains a square, there are $\psi(q)=q\prod_{p|q}(1+\frac{1}{p})$  points on the projective line $P_1(\mathbb{Z}_q)$ (in the Dedekind finction $\psi(q)$, the product is taken over all primes $p$ dividing $q$) and the remaining $\sigma(q)-\psi(q)\ne 0$ independent points (with $\sigma(q)$ the sum of divisors function) is playing the role of a reference frame and possess their own modular substructure. The number theoretical properties of the modular ring $\mathbb{Z}_q$ are used to count the cardinality of the symplectic group $\mbox{Sp}(2,\mathbb{Z}_q)$ \cite{Albouy2009,Vourdas2010}. In Sec. 3, we remind the established results concerning the point/line geometries attached to multiple qudit systems in dimension $p^n$, that symplectic polar spaces $W_{2n-1}(p)$ of order $p$ and rank $n$ govern the commutation structure of the observables. Here, the number theoretical functions $\sigma(p^{2n-1})$ and $\psi(p^{2n-1})$ are found to count the number of observables in the symplectic polar space and in the {\it punctured} polar space, respectively. In Sec. 4, we study composite systems when at least one of the factors $q_i$ of the Hilbert space dimension is a square. It is shown, that the non-modularity leads to a natural splitting of the Pauli graph/geometry into several copies of basic structures such as polar spaces, punctured polar spaces and related hyperdimensional structures. 

\section{Pauli graph/geometry of a single qudit}

A single qudit is defined by a Weyl pair $(X,Z)$ of {\it shift} and {\it clock} cyclic operators satisfying 
\begin{equation}
ZX-\omega XZ =0,
\label{Weylpair}
\end{equation}
where $\omega=\exp \frac {2i\pi}{q}$ is a primitive $q$-th root of unity and $0$ is the null $q$-dimensional matrix. In the standard computational basis $\{\left|s\right\rangle, s \in {\mathbb{Z}_q}\}$, the explicit form of the pair is as follows
\begin{equation}
X=\left(\begin{array}{ccccc} 0 &0 &\ldots &0& 1 \\1 & 0  &\ldots & 0&0 \\. & . & \ldots &.& . \\. & . & \ldots &.& . \\0& 0 &\ldots &1 & 0\\ \end{array}\right),~~ Z= \mbox{diag}(1,\omega,\omega^2,\ldots,\omega^{q-1}).
\label{Paulis}
\end{equation}

The Weyl pair generates the single qudit Pauli group $\mathcal{P}_q=\left\langle X,Z\right\rangle$, of order $q^3$, where each element may be written in a unique way as $\omega^aX^bZ^c$, with $a,b,c \in \mathbb{Z}_q$ \cite{Planat2007}-\cite{Planat2009}.

It will be shown in this section that the study of commutation relations in a arbitrary single qudit system may be based on the study of symplectic modules over the modular ring $\mathbb{Z}_q^2$, and conversely that the elegant number theoretical relations underlying the {\it isotropic} lines of $\mathbb{Z}_q^2$ have their counterpart in the maximal commuting sets of a qudit system. Our results may be found in various disguises in several publications where the proofs are given \cite{Havlicek2008,Albouy2009,Vourdas2010}.

Let us start with the Weyl pair property (\ref{Weylpair}) and write the group theoretical commutator as $\left[X,Z\right]=XZX^{-1}Z^{-1}=\omega^{-1} I_q$ (where $I_q$ is the $q$-dimensional identity matrix), so that one gets the expression  
\begin{equation}
\left[\omega^aX^bZ^c,\omega^{a'}X^{b'}Z^{c'}\right]=\omega^{cb'-c'b}I_q,
\label{transfer}
\end{equation}
meaning that two elements of $\mathcal{P}_q$ commute if only if the determinant $\Delta=\mbox{det}\left(\begin{array}{cc} b' &b \\c'& c \\ \end{array}\right)$ vanishes.
Two vectors such that their symplectic inner product $\left[(b',c').(b,c)\right] =\Delta=b'c-bc'$ vanishes are called perpendicular. Thus, from (\ref{transfer}), one can transfer the study of commutation relations within the group $\mathcal{P}_q$ to the study of perpendicularity of vectors in the ring $\mathbb{Z}_q^2$ \cite{Havlicek2008}.

From (\ref{transfer}), one gets the important result that the set $\mathcal{P}_q'$ of commutators (also called the derived subgroup) and the center $Z(\mathcal{P}_q)$ of the Pauli group $\mathcal{P}_q$ are identical, and one is led to the isomorphism
\begin{equation}
(\mathcal{P}_q/Z(\mathcal{P}_q),\times) \cong (\mathbb{Z}_q^2,+),
\label{iso}
\end{equation}
i.e. multiplication of observables taken in the central quotient $\mathcal{P}_q/Z(\mathcal{P}_q)$ transfers to the algebra of vectors in the $\mathbb{Z}_q$-module $\mathbb{Z}_q^2$ endowed with the symplectic inner product \lq\lq .".

\subsection*{Isotropic lines of the lattice $\mathbb{Z}_q^2$}

Let us now define a {\it isotropic line} as a set of $q$ points on the lattice $\mathbb{Z}_q^2$ such that the symplectic product of any two of them is $0 (\mbox{mod}~ q)$. From (\ref{iso}), to such an isotropic line corresponds a maximal commuting set in  $\mathcal{P}_q/Z(\mathcal{P}_q)$.

Taking the prime power decomposition of the Hilbert space dimension as $q=\prod_i p_i^{s_i}$, it is shown in (18) of \cite{Albouy2009} that the number of isotropic lines of the lattice $\mathbb{Z}_q^2$ reads 
\begin{equation}
\eta(q)=\prod_i \frac{p_i^{s_i+1}-1}{p_i-1}\equiv \sigma (q),
\label{divisor1}
\end{equation}
where $\sigma(q)$ denotes the sum of divisor function. 

\subsection*{The projective line $P_1(\mathbb{Z}_q)$}

As shown in \cite{Albouy2009}, a isotropic line of $\mathbb{Z}_q^2$ corresponds to a {\it Lagrangian submodule}, i.e. a maximal module such that the perpendicular module $M^{\perp}=M$. Let us now specialize to Lagrangian submodules that are {\it free cyclic submodules}
\begin{equation}
\mathbb{Z}_q(b,c)=\left\{(ub,uc)|u \in \mathbb{Z}_q\right\},
\label{module}
\end{equation}
for which the application $u \rightarrow (ub,uc)$ is injective. Not all Lagrangian submodules are free cyclic submodules. A point $x=(b,c)$ such that $\mathbb{Z}_q(b,c)$ is free is called an {\it admissible point}, and the set of admissible points is called the projective line
\begin{equation}
\mathbb{P}_1(\mathbb{Z}_q)=\left\{\mathbb{Z}_q(b,c)|(b,c) ~\mbox{is}~\mbox{admissible}\right\}.
\label{proj1}
\end{equation}
Following theorem 5 in \cite{Havlicek2008}, the number of points of the projective line is
\begin{equation}
|\mathbb{P}_1(\mathbb{Z}_q)|=\prod_i (p_i^{s_i}+p_i^{s_i-1})\equiv \psi(q),
\label{proj1card}
\end{equation}
where $\psi(q)=q \prod_{p|q}(1+\frac{1}{p})$ and the product is taken over all primes $p$ dividing $q$.  The proof is easy to establish since $\psi(q)$ is a multiplicative function. Note that one has $\psi(q) \le \sigma(q)$, where the equality holds if $q$ is square-free integer.

Let us give a few single qudit decompositions $q=4,8,9,12,16$ and $18$, that contain a square, one gets $\psi(4)=4(1+\frac{1}{2})=6,~ \psi(8)=8(1+\frac{1}{2})=12,~ \psi(9)=12,~\psi(12)=24,~\psi(16)=24,~\psi(18)=36$ (see table 1, column 3 in \cite{PlanatJPA}).

\subsection*{The Pauli graph of a qudit}

In the previous subsections, we investigated the bijection between sets of operators of the Pauli group $\mathcal{P}_q$ and vectors defined over the modular ring $\mathbb{Z}_q$. More precisely, from (\ref{iso}), elements of the central quotient of the Pauli group $\mathcal{P}_q/Z(\mathcal{P}_q)$ were  mapped to vectors of the lattice $\mathbb{Z}_q^2$ and, from (\ref{divisor1}) the $\sigma(q)$ isotropic lines of $\mathbb{Z}_q^2$ were mapped to its maximal commuting sets. 

One can see these bijections in a clearer way by defining the Pauli graph $\mathcal{G}_q$ of the qudit system. The Pauli graph $\mathcal{G}_q$ is constructed by taking the observables as vertices and a edge joining two commuting observables. A maximal set of mutually commuting observables corresponds to a maximum clique of $\mathcal{G}_q$, and one further defines a point/line incidence geometry with observables as points and maximum cliques as lines. One  characterizes this geometry by creating a dual graph $\mathcal{G}_q^{\star}$ such that the vertices are the cliques and a edge joins two non-intersecting cliques. The connected component of $\mathcal{G}_q^{\star}$ corresponds to the graph of the projective line $\mathbb{P}_1(\mathbb{Z}_q)$ (as  defined in previous papers \cite{Planat2007}-\cite{Havlicek2009}).

In the subsequent sections, we shall also introduce the graph $\mathcal{G}_q^{(k)}$, in which the vertices are the maximum cliques of the Pauli graph $\mathcal{G}_q$ and a edge joins two maximum cliques intersecting at $k$ points.

\subsubsection*{The quartit system}

For the four-level system, there are $4^2-1$ observables/vertices in the Pauli graph $\mathcal{G}_4$. The $\sigma(4)=7$ maximum cliques 
\begin{eqnarray}
&cl:=\{(X^2,Z^2,Z^2X^2),(X,X^2,X^3),(X^2,Z^2X,Z^2X^3),\nonumber \\
&(Z,Z^2,Z^3),(ZX,Z^2X^2,Z^3 X^3),(Z X^2,Z^2,Z^3X^2),\nonumber \\
&(ZX^3,Z^2X^2,Z^3X)\}
\label{quartit1}
\end{eqnarray}
are mapped to the following isotropic lines of $\mathbb{Z}_4^2$
\begin{eqnarray}
&il:=\{\{(0,2),(2,0),(2,2)\},\{(0,1),(0,2),(0,3)\},\nonumber \\
&\{(0,2),(2,1),(2,3)\},\{(1,0),(2,0),(3,0)\},\nonumber \\
&\{(1,1),(2,2),(3,3)\},\{(1,2),(2,0),(3,2)\},\nonumber \\
&\{(1,3),(2,2),(3,1)\}\}.
\label{quartit2}
\end{eqnarray}

From the latter list, one easily observes that non-admissible vectors belong to the first line $\{(0,2),(2,0),(2,2)\}$, that corresponds to the maximum clique $(X^2,Z^2,Z^2X^2)$. The remaing  vectors in $\mathbb{Z}_4^2$ generate free cyclic submodules of the form (\ref{module}).

The sequence of degrees in $\mathcal{G}_q^{\star}$ is obtained as $(1,0,0,0,6)$, meaning that the first clique given in (\ref{quartit1}) (of degree $0$) intersects all the remaing ones, and that cliques number $2$ to $7$ in (\ref{quartit1}) (of degrees $4$) form the projective line $\mathbb{P}_1(\mathbb{Z}_4)$. Indeed, one has $|\mathbb{P}_1(\mathbb{Z}_4)|=\psi(4)=6$. There are $J_2(4)=\phi(4)\psi(4)=12$ admissible points. 

The graph $\mathcal{G}_4^{\star}$ is strongly regular, with spectrum $\{4^1,0^{3+1},-2^2\}$ (in the notations of \cite{Planat2007}); the notation $0^{3+1}$ in the spectrum means that $0^3$ belongs to the projective line subgraph and there exists an extra $0$ eigenvalue in the spectrum of $\mathcal{G}_4^{\star}$. The automorphism group of $\mathbb{P}_1(\mathbb{Z}_4)$ is found to be the direct product $G_{48}=\mathbb{Z}_2 \times S_4$ (where $S_4$ is the four-letter symmetric group).

\subsubsection*{The $12$-dit system}
The main results for all qudit systems with $4 \le q \le 18$, such that $q$ contains a square, are given in table 1 of \cite{PlanatJPA}. We take the composite dimension $q=2^2 \times 3$ as our second illustration. There are $12^2-1=143$ observables in the Pauli graph $\mathcal{G}_{12}$. There are $\sigma(12)=28$ maximum cliques in $\mathcal{G}_{12}$, as expected. The sequence of degrees in the dual graph $\mathcal{G}_{12}^{\star}$ is found as $(4,0,\ldots,24)$, i.e. there are four cliques of degree $0$ and the remaining $\psi(12)=24$ ones have degree $12$.  

Owing to the composite character of the dimension, the structure of $\mathcal{G}_{12}^{\star}$ is more complex than in the quartit case, see Fig. 1 of \cite{Havlicek2008} for a picture. All four independent cliques intersect at the three vectors $(0,6),(6,0),(6,6)$, corresponding to the three observables $X^6,Z^6,X^6 Z^6$. The remaining $24$ cliques intersect at $0,1,2,3$ or $5$ points. The automorphism group of $\mathbb{P}_1(\mathbb{Z}_{12})$ is found to be $\mathbb{Z}_2^{12} \rtimes G_{144}$, with $G_{144}=A_4 \rtimes D_6$ and $G_{144}=\mbox{aut}(\mathbb{P}_1(\mathbb{Z}_{6}))$ the automorphism group of the sextit system.

\section{Pauli graph/geometry for multiple qudits}

In this section, we specialize on multiple qudits $q=p^n$, when the qudit is a p-dit (with $p$ a prime number). The multiple qudit Pauli group $\mathcal{P}_q$ is generated from the $n$-fold tensor product of Pauli operators $X$ and $Z$ [defined in (\ref{Paulis}) with $\omega=\exp(\frac{2i \pi}{p})$]. One has $|\mathcal{P}_q|=p^{2n+1}$ and the derived group $\mathcal{P}_q'$ equals the center $Z(\mathcal{P}_q)$ so that $|\mathcal{P}_q'|=p$. 

Following \cite{PlanatJPA,Saniga2007}, the observables of $\mathcal{P}_q/Z(\mathcal{P}_q)$ are seen as the elements of the $2n$-dimensional vector space $V(2n,p)$ defined over the field $\mathbb{F}_p$, and one makes use of the commutator  
\begin{equation}
[.,.]:~ V(2n,p) \times V(2n,p) \rightarrow  \mathcal{P}_q'
\end{equation}
to induce a non-singular alternating bilinear form on $V(2n,p)$, and simultaneously a symplectic form on the projective space $PG(2n-1,p)$ over $\mathbb{F}_p$ (for another approach, see \cite{Klimov2010}).

Doing this, the $|V(2n,q)|=p^{2n}$ observables of $\mathcal{P}_q/Z(\mathcal{P}_q)$ are mapped to the points of the symplectic polar space $W_{2n-1}(p)$ of cardinality \cite{PlanatJPA}.

The identification of $|W_{2n-1}(p)|$ to $\sigma(p^{2n-1})$ is new in this context. It is reminiscent of (\ref{divisor1}) and has still unoticed consequences about the structure of the polar space, as explained in the sequel of the paper. For $q$-level systems (single qudits), $\sigma(q)$ and $\psi(q)$ refer to the number of isotropic lines and the number of points of the projective line, respectively (as in (\ref{divisor1}) and (\ref{proj1card})). For multiple qudits, one has $q=p^{2n-1}$ and $\sigma(q)$ and $\psi(q)$ refer to the number of points of the symplectic polar space $W_{2n-1}(p)$ and of {\it punctured polar space} $W_{2n-1}(p)'$, respectively (as in (\ref{polar1}) and (\ref{polar2})). 
\begin{equation}
|W_{2n-1}(p)|=\frac{p^{2n}-1}{p-1} \equiv \sigma(p^{2n-1}),
\label{polar1}
\end{equation}
and two elements of $[\mathcal{P}_q/Z(\mathcal{P}_q),\times]$ commute iff the corresponding points of the polar space $W_{2n-1}(p)$ are collinear.
 
A subspace of $V(2n,p)$ is called totally isotropic if the symplectic form vanishes identically on it. The polar space $W_{2n-1}(p)$ can be regarded as the space of totally isotropic subspaces of the $(2n-1)$-dimensional projective space $PG(2n-1,p)$. Such totally isotropic subspaces, also called generators $G$, have dimension $p^n-1$ and their number is 
\begin{equation}
|\Sigma(W_{2n-1}(p))|=\prod_{i=1}^n (1+p^i).
\label{gens}
\end{equation}
Let us call a spread $S$ of a vector space a set of generators partitioning its points. The size of a spread of $V(2n,p)$ is $|S|=p^{n}+1$ and one has $|V(2n,p)|-1=|S|\times|G|=(p^{n}+1)\times (p^{n}-1)=p^{2n}-1$, as expected. 

Going back to the Pauli observables, a generator $G$ corresponds to a maximal commuting set and a spread $S$ corresponds to a maximum (and complete) set of disjoint maximal commuting sets. Two generators in a spread are mutually disjoint and the corresponding maximal commuting sets are mutually unbiased \cite{Planat2007}.

Let us define the punctured polar space $W_{2n-1}(p)'$ as the polar space $W_{2n-1}(p)$ minus a point $u$ and all the totally isotropic spaces passing though it \footnote{In the graph context the symbol ' means a puncture in the graph. It is not the same symbol as in the derived subgroup $G'$ of the group $G$.}.  Then, one gets
\begin{equation}
|W_{2n-1}(p)'|=\sigma(p^{2n-1})-\sigma(p^{2n-3})=\psi(p^{2n-1}),
\label{polar2}
\end{equation}
where $\sigma(p^{2n-3})$ is the size of the removed part and $\psi(q)$ is the Dedekind psi function.

\subsection*{The Pauli graph of a multiple qudit}

The symmetries carried by multiple qudit systems may also be studied with Pauli graphs. We define the Pauli graph $\mathcal{G}_{p^n}$ of a multiple $p^n$-dit, as we did for the single qudit case, by taking the observables as vertices and a edge joining two commuting observables. A dual graph $\mathcal{G}^{\star}_{p^n}$ is such that the vertices are the maximum cliques and a edge joins two non-interesting cliques. One denotes $\mathcal{G}'^{\star}_{p^n}$ the corresponding graph attached to the punctured polar space. 

Actual calculations have been performed for two- and three-qubits, and for two- and three-qutrits. Main results are in table 2 of \cite{PlanatJPA}. Here, we detail results concerning the two- and three-qubit systems. 

\subsection*{The two-qubit system}

As already emphasized in \cite{Planat2007,Saniga2007}, the two-qubit system \lq\lq is" the symplectic polar space $W_3(2)$ [i.e. $p=n=2$ in (\ref{polar1})], alias the generalized quadrangle $GQ(2,2)$, also called {\it doily}, with $15$ points and, dually, $15$ lines (see Fig. 6 in \cite{Planat2007}). One denotes the corresponding Pauli graph as $\mathcal{G}_{2^2}$. The maximum cliques are as follows
\begin{eqnarray}
& cl:=\{ 
  \{ IX,XI,XX \},\{ IX,YI,YX \},\{ IX,ZI,ZX \},  \nonumber \\
& \{ IY,XI,XY \},\{ IY,YI,YY \},\{ IY,ZI,ZY \}, \nonumber \\ 
& \{ IZ,XI,XZ \},\{ IZ,YI,YZ \},\{ IZ,ZI,ZZ \}, \nonumber \\
& \{ XY,YX,ZZ \},\{ XY,YZ,ZX \},\{ XZ,YX,ZY \}, \nonumber \\
& \{ XZ,YY,ZX \},\{ XX,YY,ZZ \},\{ XX,YZ,ZY \}\}  
\label{clGQ22}
\end{eqnarray}
where a notation such as $IX$ means the tensor product of $I$ and $X$.

The spectrum of the (strongly regular) Pauli graph $\mathcal{G}_{2^2}$ is $\{6^2,1^9,-3^5\}$ and the automorphism group is the symmetric group $\mbox{Sp}(4,2)=S_6$.

Following definition (\ref{polar2}), ones defines the punctured polar space $W_3(2)'\equiv GQ(2,2)'$ by removing a point in $GQ(2,2)$ as well as the totally isotropic subspaces/maximum cliques passing through it
[for the selected point $u\equiv IX$, the removed cliques are numbered 1 to 3 in (\ref{clGQ22})]. The punctured Pauli graph $\mathcal{G}_{2^2}'^*$ is as follows

\begin{eqnarray}
&GQ(2,2)'\Rightarrow \mathcal{G}'^*_{2^2}:\nonumber \\
&\mbox{spec}:= \{6^1,2^3,0^2,-2^6\},\nonumber \\
&\mbox{aut}(\mathcal{G}'^*_{2^2}):=G_{48}=\mathbb{Z}_2 \times S_4.
\label{dGQ22}
\end{eqnarray}
The automorphism group of the graph $\mathcal{G}'^*_{2^2}$ is similar to the automorphism group obtained from the graph of the projective line $\mathbb{P}_1(\mathbb{Z}_4)$, associated to the quartit system, although the spectrum and the commutation structure are indeed not the same. In a next paper, it will be shown that both graphs are topologically equivalent to the hollow sphere.

It is already mentioned in Sec. 3 of \cite{Planat2007} that the Pauli graph $\mathcal{G}_{2^2}$ can be regarded as $\hat{L}(K_6)$ (it is isomorphic to the line graph of the complete graph $K_6$ with six vertices). Similarly, defining $K_{222}$ as the complete tripartite graph (alias the $3$-cocktail party graph, or octahedral graph), one gets $\mathcal{G}'_{2^2}=\hat{L}(K_{222})$.

\subsection*{The three-qubit system}

For three qubits, the structure of commutation relations is that of the polar space $W_5(2)$ with $\sigma(2^5)=63$ elements and $(1+2)(1+2^2)(1+2^3)=135$ generators. The (regular) Pauli graph $\mathcal{G}_{2^3}$ has spectrum $\{30^1,3^{35},-5^{27}\}$ and $\mbox{aut}(\mathcal{G}_{2^3})=\mbox{Sp}(6,2)\cong W'(E_7)$, of order $1451520$. Two maximum cliques intersect at $0$, $1$ or $3$ points. The dual Pauli graph $\mathcal{G}_{2^3}^{\star}$ has spectrum $\{64^1,4^{84},-8^{50}\}$ and $\mbox{aut}(\mathcal{G}_{2^3}^{\star})= O^+(8,2)$. Note that $O^+(8,2)$ is related to the Weyl group of $E_8$ by the isomorphism $W(E_8)\cong \mathbb{Z}_2 . O^+(8,2)$. 

The three-qubit system is very peculiar among multiple qubit systems, having $O^+(8,2)$ as the automorphism group attached to the maximum cliques, instead of the symplectic group $\mbox{Sp}(6,2)$. 

One defines the punctured polar space $W_5(2)'$ with $|W_5(2)'|=\psi(2^5)=48$ points and the corresponding graph $\mathcal{G}'^*_{2^3}$ as follows
\begin{eqnarray}
&W_5(2)'\Rightarrow \mathcal{G}'^*_{2^3}:\nonumber \\
&\mbox{spec}:= \{56^1,4^{70},-4^{14},-8^{35}\},\nonumber \\
&\mbox{aut}(\mathcal{G}'^*_{2^3}):=\mathbb{Z}_2^6 \rtimes A_8,
\label{dW52}
\end{eqnarray}
with $A_8$ the eight letter alternating group.

The corresponding $1$-point intersection graph of the maximum cliques has spectrum $\{56^1,14^{15},2^{35},-4^{84}\}$. As shown in Sec. 4, it occurs in the study of the $3$-qubit/qutrit system. 

Thus, the {\it number of pieces} within the automorphism group of the dual Pauli graph  $\mathcal{G}^{\star}_{2^3}$ is twice the number $135$ of maximum cliques, instead of the cardinality $\sigma(p^{2n-1})$ of the symplectic polar space $W_{2n-1}(p)$, that is given in (\ref{polar1}).

To conclude this subsection, let us mention that the Weyl group $W(E_6)$ arises as the symmetry group of a subgeometry of the polar space $W_5(2)$, namely in the generalized quadrangle $GQ(2,4)$ \cite{Levay2009}. Taking the $27$ three-qubit observables shown in Fig. 3 of \cite{Levay2009}, one attaches to such a geometry a Pauli graph, that we denote $\mathcal{G}_{27}$. One gets $45$ maximum cliques of size $3$, the spectrum is $\{10^1,1^{20},-5^6\}$ and $\mbox{aut}(\mathcal{G}_{27})\cong W(E_6)$. By removing a point from $GQ(2,4)$ and the totally isotropic spaces passing through it, one gets the punctured generalized quadrangle $GQ(2,4)'$. The corresponding dual Pauli graph $\mathcal{G}_{27}'^*$ has spectrum $\{32^1,2^{24},-4^{20}\}$ and automorphism group $\mathbb{Z}_2 \wr A_5$ (where $\wr$ means the wreath product of groups).

The one-point intersection graph $\mathcal{G}_{45}$ of the $45$ maximum cliques is that of the generalized quadrangle $GQ(4,2)$, the dual geometry of $GQ(2,4)$. The spectrum of $\mathcal{G}_{45}$ is $\{12^1,3^{20},-3^{24}\}$ and $\mbox{aut}(\mathcal{G}_{45})\cong W(E_6)$. The automorphism group of the punctured Pauli graph $\mathcal{G}_{45}'^*$ is isomorphic to the Weyl group $W(F_4)$ of the $24$-cell. 

 This view fits the one proposed in our paper \cite{Planat2009}.

\section{Pauli graph/geometry of multiple qudit mixtures}

As before, $\mathcal{G}_q$ is the Pauli graph whose vertices are the observables and whose edges join two commuting observables. A dual graph of the Pauli graph is $\mathcal{G}^*_q$ whose vertices are the maximum cliques and whose edges join two non-intersecting cliques. In this section, we also introduces $\mathcal{G}_q^{(k)}$, the graph whose vertices are the maximum cliques and whose edges join two maximum cliques intersecting at $k$ points. 

First of all, as shown in Sec. 6 of \cite{Havlicek2007}, a qudit mixture in composite dimension $q=p_1\times p_2\times \cdots \times p_r$ ($p_i$ a prime number), identifies to a single $q$-dit. Since the ring $\mathbb{Z}_q$ is isomorphic to the direct product $\mathbb{Z}_{p_1} \times \mathbb{Z}_{p_2} \times \cdots \mathbb{Z}_{p_r}$ the commutation relations arrange as the $\sigma(q)\equiv \psi(q)$ isotropic lines of the lattice $\mathbb{Z}_q^2$, that reproduce the projective line $\mathbb{P}_1(\mathbb{Z}_q)=\mathbb{P}_1(\mathbb{Z}_{p_1}) \times \mathbb{P}_1(\mathbb{Z}_{p_2})\times \cdots \times\mathbb{P}_1(\mathbb{Z}_{p_r})$.

As an illustration, we detail the case of the two-qubit/qutrit system and the case of a two-quartit system. More examples can be found in \cite{PlanatJPA}.

\subsection*{The two-qubit/qutrit system}

The Pauli graph of the two-qubit/qutrit system contains $143$ vertices and $60$ maximum cliques. The incidence graph of the maximum cliques is found to reproduce the projective line over the ring $\mathbb{F}_4\times \mathbb{Z}_2 \times \mathbb{Z}_3$ \cite{Planat2007bis} and the spectrum of the dual Pauli graph $\mathcal{G}_{2^2 \times 3}^{\star}$ is $\{24^1,6^5,2^{27},-2^{15},-6^9,-8^3\}$. Maximum cliques of the Pauli graph intersect each other at $0$, $1$, $2$ or $5$ points. In $\mathcal{G}_{2^2 \times 3}^{\star}$, there are $480$ maximum cliques of size $3$ and $720$ maximum cliques of size $4$, to which one can attach the same number of non-complete sets of mutually unbiased bases. 

An interesting subgeometry of the two-qubit/qutrit system is found by taking the incidence graph $\mathcal{G}_{2^2 \times 3}^{(5)}$ of maximum cliques of the Pauli graph intersecting each other at $5$ points. The spectrum of this graph is $\{6^1,1^9,-3^5\}^4$ corresponding to four {\it copies} of the doily $GQ(2,2)$ [alias $\hat{L}(K_6)$]. The automorphism group of this geometry is $S_6^4 \rtimes S_4$. Similarly, the spectrum of the incidence graph for maximum cliques intersecting at two points is $\{8^1,2^{5},-2^{9}\}^4$, that represents four copies of the triangular graph $L(K_6)$. Thus, the doily is a {\it basic constituent} of the two-qubit/qutrit system and builds up its commutation structure, as one may have expected.

\subsection*{The two-quartit  system}

The two-quartit system corresponds to the decomposition $q=4 \times 4$ of the Hilbert space dimension. The Pauli graph $\mathcal{G}_{4 \times 4}$ contains $151$ maximum cliques. The connected subgraph $\mathcal{G}_{4 \times 4}^{\star (c)}$ of the dual graph $\mathcal{G}_{4 \times 4}^{\star}$ corresponds to $120$ maximum cliques of the Pauli graph that intersect each other at $0$, $1$, $3$ or $7$ points. The graph $\mathcal{G}_{4 \times 4}^{(7)}$ featuring the intersection of the $120$ maximum cliques at $7$ points has spectrum $\{-3^1,3^1,-1^3,1^3\}^{15}$, that corresponds to $15$ copies of the cube graph. The automorphism group $G_{48}$ of the cube graph is similar to that of the punctured generalized quadrangle $GQ(2,2)'$. The automorphism group of the selected geometry $\mathcal{G}_{4 \times 4}^{(7)}$ is found to be $G_{48}^{15} \rtimes S_{15}$.

The remaining $31$ cliques intersect each other at $3$ or $7$ points. The $3$-clique intersection graph still splits into a isolated clique and a connected component of $30$ maximum cliques. The connected component, of spectrum $\{28^1,0^{15},-2^{14}\}$, is the $15$-cocktail party graph, i.e. the dual graph of the $15$-hypercube graph.

\section{Conclusion}

Number theoretical functions $\sigma(q)$ and $\psi(q)$ enter into the structure of commutation relations of Pauli graphs and geometries. Theu are also related to the Riemann hypothesis, as described in \cite{PlanatSole2010}. More precisely, there is the Robin criterion

$$\frac{\sigma(q)}{q \log \log q}-e^{\gamma}<0 ~~\mbox{for}~~\mbox{any}~~ q \ge 5041$$

and the criterion derived in \cite{PlanatSole2010}
\begin{equation}
\frac{\psi(N_q)}{N_q \log \log N_q}-\frac{e^{\gamma}}{\zeta(2)}>0 ~\mbox{for}~~\mbox{any}~~ q ~~\ge 31,
\label{KMS}
\end{equation}
where $N_q=\prod _{k=1}^q p_k$ is a primorial number (the product of the first $q$ primes), $\gamma=\mbox{lim}_{n \rightarrow \infty}\sum_{k=1}^n \frac{1}{k}-\log n \normalfont\approx 0.577$ is the Euler-Mascheroni constant and $\zeta(2)=\pi^2/6$. Note that an equation similar to (\ref{KMS}) is obeyed by the expectation value of low temperature phase states $\mbox{KMS}_{\beta}(e_q)=q^{-\beta} \prod_{p|q} \frac{1-p^{\beta -1}}{1-p^{-1}}$ in Bost and Connes theory \cite{PlanatKMS}. The case (\ref{KMS}) corresponds to the integer temperature $\beta=3$.

The structural role of symplectic groups $Sp(2n,p)$ has been found, as expected. Other important symmetry groups are $G_{48}=\mathbb{Z}_2\times S_4$, $G_{144}=A_4 \times D_6$ and $W(E_6)$. The group $G_{48}$ is first of all the automorphism group of the single qudit Pauli group $\mathcal{P}_1$ and is important in understanding the CPT symmetry \cite{PlanatCPT}. In this paper, it arises as the symmetry group of the quartit, of the punctured generalized quadrangle $GQ(2,2)'$ (see  \ref{dGQ22})) and as a normal subgroup of many systems of qudits. The torus group $G_{144}$ occurs in the symmetries of the $6$-dit, $12$-dit, $18$-dit and $24$-dit systems. The Weyl group $W(E_6)$ happens to be central in the symmetries of three-qubit and multiple qutrit systems. The understanding of symmetries in the Hilbert space is important for the applications in quantum information processing.

\end{document}